**Title:**

# Radiation damage to normal mammalian tissue *in vivo* with laser-driven protons at ultra-high instantaneous dose rate

**Authors**

Lieselotte Obst-Huebl[1,*,†] (lobsthuebl@lbl.gov)
Jamie L. Inman[1,†] (jlinman@lbl.gov)
Jared De Chant[1] (jtdechant@lbl.gov)
Kei Nakamura[1] (knakamura@lbl.gov)
Sahel Hakimi[1] (shakimi@avalanche.energy)
Morgan Cole[1] (morgan.cole@yale.edu)
Hang Chang[1] (hchang@lbl.gov)
Cameron G. R. Geddes[1] (cgrgeddes@lbl.gov)
Anthony J. Gonsalves[1] (ajgonsalves@lbl.gov)
Jian-Hua Mao[1] (jhmao@lbl.gov)
Carl B. Schroeder[1] (cbschroeder@lbl.gov)
Blake A. Simmons[1] (basimmons@lbl.gov)
Jeroen van Tilborg[1] (jvantilborg@lbl.gov)
Eric Esarey[1] (ehesarey@lbl.gov)
Antoine M. Snijders[1,‡] (snijders1@llnl.gov)

**Affiliations**

[1] Lawrence Berkeley National Laboratory, 1 Cyclotron Rd, Berkeley, CA 94720.
*Corresponding Author, lobsthuebl@lbl.gov.
[†]These authors contributed equally to this work.
[‡] Present address: Lawrence Livermore National Laboratory, 7000 East Ave., Livermore, CA 04550.

**Funding:**

LaserNetUS (lasernetus.org), Offices of Fusion Energy Sciences, and High Energy Physics, of the Office of Science of the U.S. Department of Energy Contract No. DE-AC02-05CH11231.
Lawrence Berkeley National Laboratory Laboratory Directed Research and Development (LDRD) funding provided by the Director.

**Competing interests:** Authors declare that they have no competing interests.

**Data and materials availability:** All data are available in the main text, the supplementary materials, and in the repository doi.org/10.5281/zenodo.18263365.

## Abstract

The differential sparing of normal tissues relative to tumor control observed at ultra-high dose rates, referred to as the FLASH effect, has recently gained considerable attention. The therapeutic advantages of FLASH radiotherapy are expected to be further amplified through the use of protons and ions, which enable precise dose deposition at tumor depth while minimizing irradiation of healthy tissues proximal and distal to the target. Nevertheless, the mechanism underlying this sparing effect remains poorly understood. Laser-driven proton accelerators are capable of delivering uniquely high instantaneous dose rates in ultrashort bunches. Here, we report the first *in vivo* investigation of normal tissue response to laser-driven proton irradiation, with exposures to 8 MeV protons, delivering total doses up to 51 Gy at 2 Gy per laser shot. Our findings reveal a reduction in tissue swelling following laser-driven proton treatment compared with X-ray irradiations at conventional dose rates. RNA sequencing identified differential gene expression associated with immune and epidermal programs following laser-driven proton irradiations at two different dose levels.

## Teaser

This study defines healthy-tissue responses to laser-driven proton irradiation, informing the development of next-generation radiotherapy.

## Introduction

Laser-driven (LD) proton sources[1,2] are an emerging technology enabling multi-disciplinary research with societal impact in areas like radiobiology[3,4], radioisotope production[5], material science[6,7], laboratory astrophysics[8], and fusion energy[9]. These compact sources produce proton bunches of unique parameters, namely a high flux of up to $10^{12}$ particles per bunch[10], short picosecond scale duration at the source, and low transverse emittance[11], using modern, high-power table-top laser technology on a small footprint. Particularly the high peak current within a LD proton bunch sets these sources apart from what can typically be achieved with conventional radiofrequency-driven accelerator technology. This is of interest for research into the FLASH effect that requires radiation sources capable of ultra-high dose rate and has received increasing attention in the past 10 years for its potential to revolutionize cancer radiation therapy[12,13].

The FLASH effect describes the differential sparing of normal tissue from adverse effects following radiation exposure, while maintaining an equivalent tumor response, when exposing tissue to radiation of ultra-high dose rate (UHDR) of >40 Gy/s in sub-second exposures[14,15]. This striking normal-tissue sparing raises the possibility of widening the therapeutic window in radiation oncology - enabling escalation of tumor dose per fraction while reducing late toxicities, such as fibrosis and secondary malignancies[16,17]. Consequently, the FLASH effect has become a major focus of preclinical and early clinical research, with a rapidly growing body of *in vitro*, *in vivo*, and even human studies[15,18–20]. While the overall body of research shows a promising picture of normal tissue sparing and consistent tumor killing over many normal tissues and cancers, intriguingly, there is still a lack of understanding regarding the underlying mechanisms of the FLASH effect, as well as the key radiation properties to trigger it[21]. Notably, most studies to date have employed electron beams with energies up to 10 MeV and tunable dose rate with peak dose rates up to $10^7$ Gy/s[18].

Since their first experimental demonstration, LD proton sources have been discussed as a promising new technology for proton radiotherapy, due to their potential of a reduced footprint and lower cost, and hence broader accessibility[22–24]. While the maximum achievable proton energies have increased to >100 MeV in experiments[25–27], they currently remain too low to penetrate deep enough into the human body to warrant consideration of a technology transfer to the clinic[28]. Maybe even more importantly, reliable turn-key operation of these sources is yet to be demonstrated. However, even at lower energies, laser-proton sources have been used for a variety of radiobiological studies[3,29–36]. Several review articles about the FLASH effect have highlighted the potential of LD protons for FLASH radiobiology in the short term and FLASH radiotherapy in the longer term[13,18,37,38]. The ultra-high instantaneous dose rate (UHIDR) of up to $10^9$ Gy/s within a single proton bunch[12] is orders of magnitude higher than the dose rate that can typically be achieved with FLASH-modified proton machines, which currently reach up ~$10^4$ Gy/s instantaneous dose rate[18]. To achieve FLASH dose rates, these machines often need to focus the proton beam to ≲100 μm diameter areas, then requiring rastering (spatial fractionation) to irradiate laterally extended samples. This poses challenges in isolating the FLASH effect, as micro- or mini beam therapy has shown its own advantages at conventional dose rates[39]. In contrast, LD proton exposure setups have been established that deliver UHIDR to larger irradiation fields of up to 10 mm diameter[4,40–43].

Currently, the LD proton mean dose rate is limited by the drive laser's repetition rate, which is 1-10 Hz with today's multi-Joule laser systems. Therefore, in most cases where the total dose applied per shot is below 40 Gy, the mean dose rate as well as the total duration of irradiation would not fall into the nominal limits of FLASH as defined in the literature. However, sparing effects have been observed when operating outside these limits[44–46], and it is not well understood and the topic of ongoing research to elucidate how different radiation parameters, including instantaneous and mean dose rate, affect the occurrence of the FLASH sparing[21]. Additionally, more research is needed to explore the influence of different beam depth dose profiles, including a higher linear energy transfer (LET) in the case of protons and ions, as well as other proton and ion-specific parameters, such as the rapid dose fall-off at the distal edge of the Bragg peak[47]. LD protons, which are tunable in energy without the need for degraders and are accompanied by heavier ions such as carbon and oxygen that are also produced in the same laser shot, represent a unique beam modality to explore the higher LET FLASH space in the UHIDR regime.

Lasers that are capable of driving LD sources are available on the industrial, national laboratory, and university scale and are therefore more easily accessible than FLASH-enabled conventional proton accelerators[19,48]. They represent an important resource to complement existing proton facilities in exploring a unique parameter range in UHIDR research. Previous studies have explored normal tissue sparing with LD protons *in vitro*[4,33,35,43,49,50] and with zebrafish embryos *in vivo*[36,43,51]. Here we present the first study of radiation damage to mammalian normal skin tissue *in vivo* following exposures to LD proton beams at UHIDR. This work represents an essential step towards establishing FLASH research with LD protons[21] and provides radiobiological data addressing a major clinical challenge: skin toxicity, which affects up to 90% of patients receiving radiation therapy because the skin inevitably receives an entrance dose during treatment of deep-seated tumors[52]. We also explore the effect of dose fractionation in this unique UHIDR proton regime, by spreading the total dose application over multiple subsequent days. Fractionation is an established method in conventional radiotherapy to reduce normal tissue toxicity[52].

We assessed acute skin toxicity and radiation-induced swelling in the ears of female BALB/c mice following single and fractionated LD proton exposures targeting a total dose of 40 Gy, generated and transported to the sample site in a compact irradiation setup at the Berkeley Lab Laser Accelerator petawatt (BELLA PW) laser system[53]. Due to day-to-day fluctuations in proton beam performance affecting the real-time dose estimate used to guide irradiation, delivered total doses across the four exposure groups ranged from 36.5 Gy to 51.0 Gy. Thin murine ears provide an ideal model for proton radiobiology, as they are fully penetrated by the 8 MeV proton beams used here, while allowing the rest of the body to be geometrically shielded. The 40 Gy target dose was selected based on prior BALB/c mouse ear swelling results[54], which showed substantial swelling at corresponding X-ray doses without inducing morbidity that would preclude sample analysis. We investigated ear thickness, erythema, and desquamation, that served as direct measures of acute radiation response[54], and that were recorded up to 85 days post-irradiation. We also performed RNA-sequencing at multiple time points to characterize immune and epidermal transcriptional programs.

Radiation effects appeared similar across single- and multi-fraction LD proton exposures. Treatment with a delivered single-fraction of 36.5 Gy LD protons showed reduced tissue effects compared with conventional X-rays where the applied dose was adjusted conservatively with proton radiobiological effectiveness values from the literature, suggesting a possible relative sparing. RNA-seq analysis showed gene expression changes consistent with upregulation of genes associated with repair programs at moderate LD proton doses, and consistent with their suppression at higher doses. This may indicate a dose-dependent trend, where tissue transitions from successful healing to broad functional exhaustion.

## Results

### BELLA proton beamline for 8 MeV irradiations

An LD proton source and magnetic transport beamline were implemented at the BELLA PW lasers' short focal length setup, called *iP2*[55,56] (interaction point 2), to enable irradiations of murine ear samples with 8 mm diameter proton beams. Figure 1a displays the schematic setup of the beamline at the BELLA PW. The proton bunches were generated via target normal sheath acceleration[57] (TNSA) from laser pulses interacting with a 13 μm thick Kapton foil spooled on a motorized tape drive target assembly. The BELLA PW laser system was used to deliver pulses with 7 J pulse energy, 60 fs pulse length and approximately 10—30 s pulse separation (shot on demand) to the tape drive target at an incidence angle of 30°. To minimize shot-to-shot fluctuations of the proton charge recorded at the biological sample (Fig. 1b), the target was deliberately positioned at a 50 μm distance from its optimum position for the highest proton signal where the laser spot size on target is approximately (13 ± 2) μm full width at half maximum (FWHM), compared to a 3 μm FWHM at best focus. While this resulted in a reduced maximum proton energy of 11-13 MeV and a factor ~2 lower charge at the sample site, it maximized the stability of charge per bunch, which was favored in this experiment. An example of beam stability as measured with proton diagnostics at the end of the beamline is displayed in Figure 1c.

A portion of the proton beam was captured and collimated with a permanent magnet quadrupole (PMQ) doublet and transported to the sample site at 2.5 meters distance from the proton source. The PMQ doublet was followed by a 10 mm diameter aperture, a 25 μm thick Kapton scatter foil to smooth the proton beam, and then a dipole magnet to deflect the protons downwards and avoid sample exposure to X- and gamma-rays potentially also created in the laser-target interaction. All three magnetic elements were permanent magnets for their known robustness in harsh laser-plasma interaction environments and their relative compactness[51,58–59]. Individual magnet elements were arranged in a Halbach array to boost the overall magnetic field strength while allowing for a large bore diameter to improve capture efficiency[60]. The magnetic transport beamline was designed with a homemade MATLAB simulation code that combines the features of a traditional map code, through the integration of COSY INFINITY[61], with a Monte Carlo (MC) based radiation transport code. It also allows for a flexible extension with additional magnetic elements to transport up to 30 MeV protons[62]. Lead shielding was implemented surrounding the proton beam axis that blocked X- and gamma-rays along the direct line of sight between the TNSA target and the sample location.

The proton beam exited the vacuum chamber system through a 25 μm thick Kapton window and reached the sample site in air at 14.5 cm after the window. Any heavier ions produced during the laser-target interaction that share the same magnetic rigidity as the transported protons would have been stopped by the upstream scatter foil and vacuum window. The off-beam axis location of the sample holder and the proton energy loss due to transmission through the foil, the window and the air gap resulted in an on-sample proton energy range of 5-9 MeV, with the peak of the distribution at 8 MeV, as derived from simulations of the beamline benchmarked with proton spectral measurements (Fig. 1d, refer to the Materials and Methods section for more details). An average linear energy transfer (LET) in 250 μm thick water samples was estimated as (6.9 ± 2.0) keV/μm. Despite the use of the dipole to deflect the protons downwards onto the samples, the energy spectrum was largely uniform across the vertical extent of the samples, due to the beam being only loosely collimated by the quadrupole assembly, rather than focused (see Supplementary Figure 1). Online and offline diagnostics were used in different locations along the beamline to characterize the beam parameters and provide online dosimetry for every sample irradiation. The dosimetry methods are described in the following section.

## Dosimetry for normal mouse ear irradiations

Sample irradiations were supported by online dosimetry to guide application of the target dose to each sample, with the final delivered dose to each sample subsequently determined with high accuracy via post-hoc calibration using RCF measurements collected on each irradiation day. An integrating current transformer[63] (ICT) measured the proton bunch charge for every laser shot right before the sample location in air. In preparation of sample irradiations, calibrated radiochromic films (EBT3, Ashland) were used to characterize the beam profile and energy spectrum at 52 mm after the TNSA source, as well as the lateral beam uniformity at the ear location. A single film was attached to the back of the sample holder for *in situ* dose measurements of the protons transmitted through the sample. While this served as the primary method for determining the total applied dose, several measurements were affected by fur strands or slight folds in the mouse ear skin. Because such obstructions would result in an underestimation of the delivered dose if based solely on RCF dosimetry, the total applied dose for these specific instances was derived from dose-calibrated ICT measurements. An additional film was placed even further downstream to monitor a higher proton energy bin. A scintillator coupled to a CCD readout was used at the sample location for initial beam tuning and was then placed at the end of the beamline for auxiliary online beam metrology during sample irradiations. In that location it served to measure protons close to the spectral cutoff. Together with the proton spectrum at the source, these diagnostics informed and benchmarked simulations of the proton transport and energy loss through the beamline. The dash-dotted grey lines in Fig. 1d correspond to the largest spectral deviations observed between separate days of irradiation.

Figure 2 shows results of the dosimetry (doses for individual sample exposures are listed in Supplementary Table 1). The beam uniformity was measured daily after beam tuning with an EBT3 film in the sample location (Fig. 2a) and showed an average lateral dose variation of 7% over the central 7 mm diameter area of the beam. The dose measured with film at the ear location served to calibrate the ICT in a dedicated dosimetry run. With the ICT thus serving as an online dosimeter, the dose applied on every shot could be recorded and analyzed in-situ. The average dose per shot over all sample exposures was (2.0 ± 0.4) Gy. Proton bunches were accumulated on the samples at approximately 1 shot every 10–30 seconds to reach the prescribed dose of 40 Gy. The online dosimetry with ICT served to guide sample irradiations and allowed us to account for source fluctuations from shot to shot within a day of irradiations, by adjusting the total number of LD proton bunches per sample (Fig. 2b). Day-to-day fluctuations in proton beam performance were not fully captured by the fixed real-time dose estimate used to guide irradiation, resulting in delivered total doses ranging from 36.5 to 51.0 Gy across groups, as determined by post-hoc calibration using RCF measurements collected on each irradiation day (see Proton dosimetry, Materials and Methods). RCF dose measurements were conducted using two EBT3 films placed behind the sample. The dose measured with the films was converted to dose on the sample with the help of energy loss simulations that took into account the proton spectrum at the sample, as well as corrections to the film sensitivity due to the protons' LET in the film[64].

Samples were exposed in four dose and fractionation groups. The table in Figure 2c displays the average total dose ($D$), sample to sample variations ($SSV$), and dose error ($\Delta D$) that samples received in each dose group. Thanks to our adaptive irradiation protocol guided by online dosimetry, the overall $SSV$, representing the average of the within-group standard deviations (A: 4%, B: 5%, C: 9%, and D: 3%), could be reduced to 5%, significantly lower than the 17% $SSV$ in our previous study[4]. The relative uncertainty on the total dose estimate over all samples and dose groups is 12%, including sources of uncertainty like the $SSV$, the lateral dose variation, fluctuations in the proton source spectrum, and uncertainties from the ICT and EBT3 calibrations. To estimate the instantaneous dose rate applied to the sample within each proton bunch we derived the proton bunch length from simulations as the time delay between the beam current rising above and falling below 1/e (36.8%) of the peak current. This revealed a bunch length of 11 ns. The fraction of the proton bunch contained within this bunch length was 72%. In combination with the dose per shot of 2 Gy this results in an instantaneous dose rate of $1.3\times10^8$ Gy/s. Table 1 summarizes the beam parameters used for sample exposures.

For a number of dosimetry shots, a Faraday Cup (FC) was used for additional charge measurements at 2597 mm from target and at a height which matched the same dispersion angle as the beam applied to the samples. The FC had a circular beam entrance opening of 38 mm and therefore captured a smaller (spatially filtered) portion of the beam than measured with the ICT, but larger than the sample diameter. A linear correlation between the ICT and FC was established, which in combination with linear correlations between the ICT and in-situ EBT3 measurements (performed on the beam size applied to the samples), and between the ICT and the light emitted from the scintillator (refer to the Materials and Methods section), provided additional evidence for the reliability of the ICT dosimetry.

## Normal ear response to laser-driven proton irradiation *in vivo*

We investigated radiation-induced damage caused by LD protons in healthy tissues *in vivo*. Acute skin toxicity and radiation-induced swelling were assessed in murine ears following high-dose exposures. RNA-sequencing at multiple time points was performed to characterize immune and epidermal transcriptional programs. Using a custom-designed holder (Fig. 6 in Materials and Methods section), the left ear pinna of female BALB/c mice was immobilized and irradiated with LD protons delivered either as a single acute fraction or fractionated over multiple days. Four exposure groups were examined: A (36.5 Gy, 1 fraction), B (39.6 Gy, 2 fractions), C (42.1 Gy, 4 fractions), and D (51.0 Gy, 1 fraction), each with an average of eight animals.

<u>Ear thickness, erythema and desquamation score</u>
Ear thickness was measured with a Mitutoyo micrometer for up to 85 days post-irradiation (after the last exposure day) and compared to a sham-irradiated cohort. Figure 3 shows individual animal measurements across all groups. Swelling typically peaked between days 20–30 post-irradiation, with substantial inter-animal variability even within the same exposure group. When averaged across animals, maximum ear thicknesses of (286 ± 28) µm (mean ± standard deviation of group A, day 24), (298 ± 35) µm (group B, day 28), (316 ± 28) µm (group C, day 21), and (468 ± 260) µm (group D, day 27) were observed. Accounting for dose uncertainties, groups A, B, and C allowed assessment of fractionation at comparable total doses. No significant protective effect of fractionation was observed in terms of reducing maximum ear swelling. The combined erythema and desquamation scores followed a similar pattern. In groups A-C, erythema and desquamation rose to a maximum score of ~0.5 around day 20, returning to baseline by day 40. In contrast, group D exhibited more severe responses: two animals developed a score of 3.0, coinciding with the largest increases in ear thickness, while the remaining animals in this group showed moderate scores of 0.5–1.0 (Supplementary Figure 2). A complete list of irradiated samples, dose fractions, and measurements of ear thickness, erythema, and desquamation is provided in Supplementary Table 1.

Two mouse cohorts were irradiated with conventional dose rate 300 kVp X-rays from an XRAD320 source. The cohorts received single-day exposures of 35.1 Gy (33.1 mGy/s, group E) and 40.8 Gy (30.9 mGy/s, group F), respectively. These dose levels were selected to provide comparison with our 36.5 Gy/1 fraction LD proton cohort at BELLA (group A), accounting for potential radiobiological effectiveness (RBE) differences, i.e., the lower X-ray dose relates to the proton dose directly, while the higher X-ray dose incorporates an RBE factor of 1.1, where $D_{x\text{-}ray} = 1.1\times D_p$. In the 40.8 Gy X-ray cohort, the maximum ear thickness reached (372 ± 66) µm (mean ± standard deviation) on day 23. This value is markedly lower than reported in a comparable study conducted by Girst *et al.*[65] using 40 Gy of 70 kVp X-rays delivered to female BALB/c mice with a 7.2×7.2 mm² beam, where ear swelling exceeded 900 µm at day >30. Instead, our results more closely align with those of Dombrowsky *et al.*[54], where fractionated 220 kVp X-ray irradiations (4×10 Gy over four consecutive days) produced an average ear thickness of (462 ± 46) µm on day 22 after the first fraction. For each of the *n* animals per group, ear thickness was averaged over days 18-28 (the window in which all animals were followed), and groups were compared on these per-animal values, reported as mean ± standard deviation across animals. Compared with 35.1 Gy X-rays, the nominally dose-matched comparison, the time-averaged thickness was lower after 36.5 Gy LD protons (279 ± 17 µm, $n = 8$, vs 314 ± 40 µm, $n = 8$), but only at the margin of conventional significance (Welch $p = 0.048$, Mann-Whitney $p = 0.050$). This was not robust to the choice of window (days 1-28: $p = 0.054$ and 0.065) or summary measure (per-animal peak thickness: $p = 0.19$ and 0.25). When groups were instead compared at each time point separately, no time point was significant after Holm correction for the number of time points. The reduction was larger and clearly significant relative to 40.8 Gy X-rays, the RBE-weighted dose (363 ± 42 µm, $n = 10$; Welch $p = 7.4\times10^{-5}$, Mann-Whitney $p = 9.1\times10^{-5}$). In the time point comparisons, 12 of 16 time points remained significant after Holm correction (Holm-adjusted $p \leq 0.05$; Fig. 3a). A linear mixed-effects model with animal as random effect gave a consistent result for the comparison with 40.8 Gy X-rays. When benchmarked against published data, our 51.0 Gy LD proton cohort (group D) exhibited significantly less maximum ear swelling than BALB/c mice irradiated with 60 Gy homogeneous field protons at low dose rates, but more swelling than when exposed to the same total dose with minibeams[65].

RNA sequencing and differential expression analysis

To define how dose and time shape tissue responses to LD proton irradiation, we performed RNA sequencing and differential gene expression (DEG) analysis on ears from select treatment groups. We first examined temporal changes in BELLA Group A by sequencing four ears collected 28 days after 36.5 Gy exposure and four ears collected 84 days after the same dose. Genes meeting a ≥1.5-fold change threshold and a Benjamini–Hochberg–adjusted *p*-value (FDR) < 0.05 were used as input for linear discriminant analysis (LDA). LDA revealed clear separation between the 28-day and 84-day irradiated transcriptomes, with each forming distinct clusters along linear discriminants 1 and 2 (LD1, LD2). In the LD1-LD2 space, LD1 predominantly captured the distinction between sham and irradiated transcriptomes, whereas LD2 captured the time-dependent divergence between the 28-day and 84-day irradiated responses. Both irradiated groups were also well separated from their corresponding sham controls, whereas the 28-day and 84-day sham controls overlapped extensively, indicating stable baseline transcriptional states over time (Fig. 4a). Supplementary Figure 4 includes the gene-set overlap (Venn diagrams) and a hierarchical-clustering heatmap demonstrating clear segregation of the treatment groups.

At 28 days post-irradiation, genes upregulated after BELLA 36.5 Gy exposure were significantly enriched for Gene Ontology (GO) Biological Processes associated with cytokine-signaling, antigen-presentation, and keratinocyte-differentiation pathways, consistent with an inflammatory and epithelial-repair transcriptional state. By 84 days, these immune-associated enrichments were largely absent, and downregulated genes were instead enriched for muscle-system and morphogenesis pathways, suggesting persistent suppression of structural programs. Notably, genes associated with keratinocyte differentiation remained over-represented among upregulated genes at 84 days, indicating ongoing epidermal turnover. In contrast, enrichment of nuclear division, chromosome-segregation, and related cell-cycle terms among upregulated genes suggested increased proliferative gene activity despite broader downregulation of structural pathways (Fig. 4b). Supplementary Figure 5 provides gene-process connectivity visualizations.

We next applied the same LDA strategy to evaluate dose- and source-dependent differences at 84 days post-exposure, again using genes with ≥1.5-fold changes and including gene-set overlap and hierarchical-clustering analyses (Supplementary Figure 6). In the LD1–LD2 space, LD1 predominantly captured the distinction between sham and irradiated transcriptomes, whereas LD2 captured variation associated with radiation dose and modality. BELLA Group A (36.5 Gy) formed a distinct cluster clearly separated from BELLA Group D (51.0 Gy), indicating dose-dependent divergence in late transcriptional states. In contrast, BELLA Group D and XRAD Group F (40.8 Gy) showed partial overlap, suggesting shared transcriptional features between high-dose proton and conventional X-ray exposures at this late time point. As expected, sham controls from all groups overlapped tightly with one another and remained well separated from all irradiated conditions (Fig. 4c).

Pathway-level comparison of enriched GO Biological Processes revealed clear modality- and dose-dependent differences in late transcriptional responses (Fig. 4d; gene–process connections in Supplementary Figure 7). In BELLA Group A, upregulated genes were enriched for keratinocyte-differentiation and cell-cycle–related terms (e.g., nuclear division), whereas downregulated genes were enriched for muscle-system, morphogenetic, and cell-structure–associated processes, patterns partially shared with the XRAD group F. XRAD Group F uniquely showed enriched antigen-presentation and MHC class I–related terms among upregulated genes and displayed enriched ion-transport and morphogenesis-related terms among downregulated genes. BELLA Group D exhibited a distinct profile in which downregulated genes were highly enriched for immune- and leukocyte-associated processes (e.g., leukocyte-mediated immunity, B-cell–mediated immunity, cell chemotaxis, regulation of immune effector processes), with relatively few enriched terms among upregulated genes at this threshold. This pattern is consistent with broad late-phase attenuation of immune and stromal gene programs at very high dose.

Hallmark pathway enrichment corroborated these distinct epithelial, immune, and transcriptional patterns across doses and modalities (Supplementary Figure 8a). Leading-edge Hallmark genes further clustered into interferon/antigen-presentation, metabolic, and stromal modules that distinguish resolving low-dose BELLA from persistent XRAD activation and high-dose BELLA suppression (Supplementary Figure 8b). Together, these data indicate that low-dose BELLA retains epithelial turnover and proliferative gene activity, XRAD engages antigen-presentation and immune programs alongside persistent structural suppression, and very high-dose BELLA produces a transcriptional profile consistent with late-phase attenuation of multiple immune and tissue-repair modules.

## Discussion

Radiobiological research into the FLASH effect has so far been dominated by studies performed using electron beams of up to 10 MeV. This is due to their broad availability, as well as straight-forward modifications to enable UHDR sample exposures. However, strongest benefits of FLASH radiotherapy are expected when combining the differential normal tissue sparing of FLASH with a targeted tumor irradiation at depth using protons and ions, where due to the inverse dose deposition in the Bragg peak, comparatively less dose is applied to the tissue in front of and behind the tumor. Harnessing the full benefits of the FLASH sparing effect in radiotherapy would require a mechanistic understanding of the effect, which has been elusive to date. Radiobiology studies into proton FLASH are emerging at a significantly slower pace, due to the limited availability of FLASH-capable proton facilities. Existing studies are typically performed at high proton energies and using the low LET plateau, because reducing the energy via beam degraders also reduces the proton current at the sample site, usually limiting the maximum dose rate achievable to below FLASH dose rates. Moreover, focusing the proton beam and rastering a sample is often necessary to achieve high proton dose rates. Therefore, isolating the effects of ultra-high dose rate from other beneficial effects reported for spatial fractionation is difficult.

In addition to a smaller footprint and reduced operation costs, LD proton sources feature an ultra-high instantaneous (bunch) dose rate (IDR) at a moderate mean dose rate (MDR), and can be delivered with a large irradiation field and broad energy range[66], motivating their contributions in a unique source parameter space to FLASH radiobiology research. Additional benefits include straight-forward variation of the proton energy (and LET) by tuning the laser energy on target, and the simultaneous production of heavier ions that can also be used for radiobiology studies[67].

We report the first irradiation study of mammalian normal tissue using exposures of murine ears with 8 MeV LD protons produced at the BELLA PW laser of Berkeley Lab. Total doses up to 51 Gy were delivered in a maximum of four fractions, over up to four days, by accumulating 2 Gy proton bunches with bunch separations of approximately 10—30 s. Ear swelling, erythema and desquamation were monitored for a maximum of 85 days post irradiation, and RNA sequencing at selected timepoints was performed to assess immune activation, inflammation, and recovery dynamics following single-fraction exposures.

Sample-to-sample variability in ear swelling within dose groups was substantial (around 10% for groups A-C and 55% for group D) and did not correlate with variations in the applied dose (ranging from a minimum of 2% in group D to a maximum of 9% in group C; groups A and B showed 4% and 5% dose variation, respectively; all values represent standard deviations). Such biological variability has also been noted in prior *in vivo* studies at both FLASH and conventional dose rates[54,65], highlighting the need for larger sample sizes to minimize the influence of inter-animal biological variation and strengthen statistical power.

We selected dose group A (36.5 Gy, single fraction) for direct comparison with conventional dose rate 300 kVp X-rays, as this total dose had previously shown substantial swelling in BALB/c mouse exposures without inducing morbidity[54], in addition to separately revealing a marked sparing effect in a mouse leg model following proton FLASH treatment delivered via pencil beam scanning[68]. Assuming an RBE factor of 1.1, we observed significantly reduced ear swelling following LD proton exposures compared to conventional X-rays. While RBE has been extensively studied to compare radiation damage across modalities with differing LET values at conventional dose rates[69-70], analogous systematic studies are lacking in the UHDR regime. Notably, an RBE of 1.1 may even underestimate the biological effectiveness of LD protons versus conventional X-rays, as studies of clonogenic cell survival comparing these two modalities report RBE values ranging between 1.2 and 1.4[30,33,71]. If such higher RBE values are considered, the equivalent X-ray dose for comparison would be greater, further increasing the observed sparing factor of LD protons. However, the biological response could still be shaped by the comparatively low proton energy and broad energy spread used in this study. Recent work has shown that low-energy proton irradiation elicits distinct DNA damage responses linked to elevated LET and proton track structure[72], and emerging evidence suggests that proton biological effectiveness depends on the energy spectrum and radiation quality of the beam, not just absorbed dose and average LET[73].

We observed no significant change in radiation damage following single- or multi-fraction LD proton exposures. Consistent with this observation, Limoli et al. reported that FLASH sparing was maintained in a hypofractionated regime with 6 MeV electrons[74]. However, with approximately eight animals per group and substantial inter-animal variability, our study was not powered to detect modest fractionation effects. Although the delivered total doses (36.5-42.1 Gy) were comparable within the dosimetric uncertainty (12%), they were not precisely matched across fractionation groups. The absence of a significant difference should therefore be read as a lack of evidence for an effect, not as evidence of equivalence, and larger cohorts will be needed to test this.

Transcriptomic and enrichment analyses revealed clear dose- and modality-dependent differences in late tissue responses to LD protons. At 36.5 Gy, genes upregulated at 28 days were enriched for interferon signaling, antigen presentation, and epithelial-repair processes, but these enrichments were largely absent by 84 days, consistent with limited long-lived injury and more complete resolution of early inflammatory programs. In contrast, 84-day XRAD samples retained strong interferon and antigen-presentation signatures and showed deeper suppression of structural pathways. High-dose BELLA (51.0 Gy) exhibited a distinct profile in which downregulated genes were significantly enriched for leukocyte, chemotactic, metabolic, and stromal pathways, with minimal enrichment among upregulated genes, consistent with a broad late-phase attenuation of immune and reparative transcriptional programs. Leading-edge Hallmark gene modules further highlighted a transition from resolving immune/epithelial programs at moderate dose to persistent activation after XRAD and to widespread suppression at higher proton dose. Taken together, these transcriptional patterns suggest that moderate-dose LD protons maintain the capacity to resolve inflammation and reestablish tissue homeostasis, whereas X-ray exposure is associated with prolonged immune activation and high-dose LD protons induce broad suppression, suggesting potential intrinsic sparing and recovery advantages of LD proton irradiation. These transcriptomic findings should be regarded as exploratory. They rest on a limited sample size per condition, which affects power and the ability to resolve smaller effects. Confirmation in larger cohorts, together with orthogonal validation (for example by histology), will be needed to better understand these differences. To unambiguously establish a FLASH sparing effect in the ultra-high IDR and low MDR regime achieved with our LD proton source, reference measurements using conventional dose-rate protons with comparable LET and energy spectrum will be essential.

Emerging evidence suggests that the instantaneous (also called “peak” or “bunch”) dose rate plays an important role in triggering the FLASH effect. This condition is inherently satisfied for LD particle sources that deliver an extremely high IDR but typically a lower MDR due to limitations in laser repetition rate. Although the BELLA PW laser with tape-drive target can, in principle, operate at 1 Hz, our first *in vivo* study at the new beamline was carried out at a reduced repetition rate of, on average, 0.05 Hz, i.e., one shot every ~10–30 seconds. This constraint arose from the need for real-time dosimetry based on the ICT data, verification that the tape had sufficiently moved to a fresh position, and to manual launching of a laser shot, in cases when additional pulses were required to reach the prescribed dose. Future studies will benefit from straightforward automation of online dosimetry coupled with direct shot-go integration into the laser control system, which will allow operation closer to the intrinsic repetition rate of the laser. This should include daily re-calibration of the ICT and continuous monitoring of additional key proton beam parameters, such as the on-shot energy spectrum, to improve real-time targeting of the prescribed dose and enable truly controlled dosing in future studies. Looking farther ahead, the advent of efficient, high power kHz laser technology[75], combined with adequate target systems such as continuous jet technology[76–78], will make it possible to achieve simultaneously ultra-high MDR and ultra-high IDR. With variable pulse separations and proton energies (by varying the laser intensity on target), and energy-variable beam transport systems[40,62], this will enable broad systematic studies of the proton FLASH effect across a wide range of MDR and LET regimes within compact LD systems available in university and industry labs.

Future work at the BELLA proton beamline will extend proton energies to approximately 30 MeV, enabling irradiation of samples several millimeters in depth. This will allow direct assessment of normal tissue sparing alongside tumor control within a single experiment, for instance by irradiating tumors implanted in murine ears or extremities and monitoring tumor growth delay in conjunction with acute and late skin responses. Similarly, *in vitro* samples contained in standard holders, such as centrifuge tubes, will be accessible at these energies. These developments will broaden the range of FLASH radiobiology experiments feasible at the BELLA PW proton beamline. In the long term, integration of a variable energy delivery system with efficient kHz laser and target technologies will permit more flexible modulation of proton beam parameters, advancing mechanistic studies of the FLASH effect and informing optimization of proton FLASH radiotherapy protocols.

## Materials and Methods

### Experimental Design

The experiment was performed using the BELLA PW laser facility at LBNL. The BELLA PW laser was the world's first 1 Hz repetition rate 1 PW Ti:Sapphire laser system based on double-chirped pulse amplification (CPA) architecture, where a cross-polarized wave (XPW) contrast enhancement system is installed in between two CPA stages, delivering pulses with a duration down to ~35 fs FWHM at 815 nm wavelength. At BELLA iP2, a 500 mm focal length off-axis parabolic mirror is used to focus the laser pulses with up to 40 J energy to a measured spot size of 3 µm FWHM diameter at best focus. For this experiment the laser was operated at 7 J, 60 fs pulse length, and defocused to a spot size of 13 µm. Kapton tape with a thickness of 13 µm was laser irradiated with a 30-degree angle of incidence. In our target assembly, the Kapton tape is spooled into a motor-controlled tape drive system, and advanced between laser shots by two DC-motors, providing a fresh wrinkle-free target surface with a position repeatability < 10 µm. This tape drive is capable of operating at a high repetition rate up to 1 Hz. However, the experiments were performed with a shot applied roughly every 10–30 seconds, to allow for live analysis of the online dosimetry diagnostic between shots and verification that the tape had sufficiently moved to a fresh position.

### Proton beam diagnostics

The proton energy spectrum and divergence at the TNSA source were characterized by exposing a stack of calibrated radiochromic films (RCF, EBT3 from Ashland) positioned at a distance of 52 mm behind the target along the target normal. The source spectrum is displayed in Figure 5a. The total proton dose applied to every sample was measured with two RCF placed behind the sample, and the dose for every shot was measured with a calibrated integrating current transformer (ICT) located in front of the sample[63]. Charged particle bunches passing through the aperture of the ICT coil induced a current that is temporarily stored in a coaxial capacitor and then delivered to a 50 Ω load, enabling signal readout with an oscilloscope. The bunch charge was then measured by integrating over the signal trace, and multiplying with a device specific calibration factor. The dosimetry protocol and dose calibration of the ICT is detailed in the Proton dosimetry section. The proton beam profile at the sample site was characterized daily by irradiating a radiochromic film instead of the sample placed in the sample holder.

A scintillator (EJ–200 from Eljen Technology) coupled to a CCD readout (Basler) was used at the sample location for initial beam tuning and was then placed at the end of the beamline for auxiliary online beam metrology during sample irradiations. Figure 5b shows an example of the scintillator signal versus the ICT-measured charge during beam tuning, with the scintillator placed at the sample location so that its light yield can be related to the dose a sample would receive there. The laser intensity on target was varied throughout this dataset, and both diagnostics are well correlated (Pearson correlation coefficient $p$ = 0.99), indicating that changes in ICT charge closely track changes in dose. When placed at the end of the beamline during sample irradiations, the scintillator served to measure protons close to the spectral cutoff. Together with the two RCF placed behind the irradiated sample, sampling different spectral ranges of the protons, these diagnostics allowed for rough assessment of the proton spectral content for each sample that was irradiated, which served to benchmark and offset the beamline simulations.

For a number of shots during the beam tuning phase a Faraday Cup (FC) was placed at 2597 mm from the target and at a height which matched the same dispersion angle as the beam applied to the samples. The FC had a circular beam entrance opening of 38 mm and therefore measured a smaller (spatially filtered) portion of the beam than measured with the ICT. A linear correlation between the beam charge measured with the ICT and FC was established (Fig. 5c), which in combination with the correlations between the ICT and in-situ RCF measurements (performed on the beam size applied to the samples), and between the ICT and light emitted by the scintillator, provided additional confidence in the ICT dosimetry.

## Proton dosimetry

The proton dose distribution was individually measured for each sample exposure using calibrated RCFs (Ashland EBT3) placed at the back of the sample holder (17 mm from sample). The films were scanned (EPSON Expression 12000XL-PH) with all image correction features turned off with a resolution of 200 dpi in transmission mode and saved as 48-bit color tiff images. Scanning was done between 7 and 13 days after irradiation to allow for stabilization of the optical density development post-irradiation. The scanner was calibrated with a NIST-calibrated transparent step wedge to convert the raw data in the red image channel to optical density (OD), which was converted to dose. The dose was then analyzed over a 7 mm diameter circle. The OD to dose calibration was acquired by exposing films in triplicate with 300 kVp X-rays from an XRAD320 source and comparing the measured optical density with the dose measured with an ionization chamber (Radcal Accu-Dose), which is recalibrated by the manufacturer on a yearly basis. The film calibration curve is provided in Supplementary Figure 9.

Converting the dose measured with film $D_{\text{film}}$ to dose applied to the sample $D_{\text{ear}}$ involved a) accounting for the variation in film sensitivity due to the linear energy transfer (LET) of protons of a certain energy versus X-rays[64], and b) the added propagation and related energy loss between the sample and the film location. For a) we calculated the energy-dependent correction factor $\eta=D_{\text{film}}/D_{\text{film,real}}$ following equation (2) in Schollmeier *et. al.*[64] and using an average LET of (11.6 ± 0.5) keV/µm. The LET for individual particles was determined from simulations using track-averaging. The LET over the full proton beam was then obtained as a spectrum-weighted average, where contributions from different proton energies were weighted according to the proton energy spectrum (i.e., by the relative number of particles at each energy), as estimated at the film location using beamline simulations. The resulting correction factor $\eta=0.73 \pm 0.01$ was found to be consistent with EBT3 under-response reported in the literature[80-81]. The validity of our LET-sensitivity correction was further assessed by comparing the integrated charge derived from EBT3 stacks with ICT measurements on the same laser shot (see Supplementary Figure 10 for details). Unfolding and integrating the proton spectra recorded by film stacks using the same analysis protocol as for sample dosimetry yielded charge values in reasonable agreement with the ICT. For b), the ratio between dose on sample versus dose on film was established on a daily basis by exposing a "fake ear" film in the sample location and comparing the dose on the fake ear film with the dose on the dosimetry film and an additional film placed further downstream for a rough spectral analysis, to benchmark the beam transport simulations and confirm the correction factor (representative examples of the corresponding dose maps are displayed in Supplementary Figure 1). The combined correction factor $f$ for a) and b) amounted to $f = D_{\text{ear}}/D_{\text{film}} = 1.45 \pm 0.1$. In some cases, dosimetry using the film placed behind the samples was affected by skin folds in the mouse ears or small amounts of mouse hair in the beam that visibly affected the proton dose profile, which was confirmed by comparing the dose profile with photographs taken of every sample. The measurement of the total dose in those cases relied on the doses measured with the calibrated ICT described in the following.

The dose applied on every shot was measured with the ICT placed in front of the sample. First, the ICT was calibrated to RCF in a dedicated dosimetry run on a day prior to sample irradiations to establish a rough charge to dose conversion factor. This calibration was used on the sample irradiation days to monitor the accumulated dose per sample and adjust the number of shots for more accurate targeting of the total dose accounting for the effect of shot to shot variations in the proton dose. For post-analysis, that calibration was refined by re-calibrating the ICT with RCF data collected on every sample irradiation day. This was to account for fluctuations in the proton beam performance from day to day. Because the real-time calibration used to guide irradiation was not updated to reflect these day to day fluctuations, the total dose targeted during irradiation (40 Gy for all groups) deviated from the true delivered dose once determined via the day-specific, post-hoc calibration, resulting in delivered doses ranging from 36.5 to 51.0 Gy across the four dose groups (Fig. 2c). The final dose reported on the samples is a combination of both ICT and film measurements. A table summarizing the dose values applied to all samples individually is provided in the Supplementary Table 1.

The dose error $\Delta D$ was calculated using Gaussian error propagation and included the following contributions: sample to sample variation (3%-9%, standard deviation), lateral proton dose variation (7%, standard deviation), the uncertainty introduced by a varying proton spectrum to the correction factors accounting for LET sensitivity reduction and converting dose on film to dose on sample (7%, maximum range), the uncertainty introduced by the RCF dose calibration (5-10%), the uncertainty introduced by shot to shot variations during ICT calibrations (1.2%, standard deviation of the mean), and the uncertainty introduced by possible film darkening due to delayed RCF scanning[82] (3-4%).

**Proton transport system and beamline simulations**

The quadrupoles were 8-azimuthal-segment Halbach arrays made of NdFeB magnets with 1.29 T remanent magnetic field and 50 mm length manufactured by Radiabeam. The entrance of the first PMQ was placed at 34 mm from the proton source and the second at 176 mm. The first (second) PMQ had an inner bore radius of 5 mm (15 mm), resulting in an on-axis magnetic field gradient of 250 T/m (67 T/m), and an angular acceptance of ± 32 mrad in the horizontal direction and ± 80 mrad in the vertical direction of the combined doublet. A 95 mm long dipole magnet with a magnetic field of 264 mT was located in the proton beam path after the PMQ doublet to deflect the protons downward from the laser direction, and lead shielding was implemented, to avoid direct irradiation of biological samples by X- or gamma-rays also created in the laser-target interaction.

The design of the proton transport system was performed with a homemade MATLAB simulation code that combines the features of a traditional map code, through the integration of COSY INFINITY[61], with a Monte Carlo (MC) based radiation transport code[62]. This code calculates the trajectories of a large number of particles and models the energy deposition into matter along the beam path and allows for the optimization of the dose delivered to the biological sample to ensure a high dose per shot and a uniform dose profile. The code first generates a large number of macroparticles ($> 10^6$) with initial parameters to match the beam emerging from the TNSA source. These particles are then propagated through a high order transfer map, generated using COSY, that describes their propagation through the various beam optics and drift spaces. A 4th order mapping or above was used to fully capture the dynamics of the LD beam, given its large energy spread and initial divergence. The code then models the propagation of the particles through the various materials along the beam path on their way to the biological sample, as well as the energy deposition inside the sample. This modeling accounts for effects of energy degradation and scattering on the beam spectrum and profile, while describing the dose deposition on the sample and a number of RCFs placed in the beam for dosimetry, based on data from the Particle Data Group[83]. Using the rough spectral measurements performed daily at the sample site, the input spectrum was adjusted to match these day-to-day changes in the proton beam performance.

## Mouse preparation

Female BALB/cJ mice were purchased from The Jackson Laboratory and allowed to acclimate at LBNL for two weeks before experimental procedures. Animals were housed in standard micro-isolator cages containing hardwood bedding (Sani Chips; P.J. Murphy Forest Products) and provided with environmental enrichment using crinkle-cut paper nesting material. Housing conditions followed a 12-hour light/dark cycle, with ambient temperature maintained at 21.9 ± 2 °C and relative humidity at 54.3 ± 10%. Mice had unrestricted access to PicoLab Rodent Diet 20 (5053) and water. Sentinel monitoring confirmed that all animals were free of common murine pathogens, including MHV, Sendai virus, PVM, *Mycoplasma pulmonis*, TMEV (GDVII), Reovirus-3, parvovirus, EDIM, LCM virus, and ectromelia virus. All animal procedures were conducted in accordance with the NIH *Guide for the Care and Use of Laboratory Animals* under PHS Assurance number D16-00031 (A3054-01)**.** Experimental protocols were approved by the LBNL Animal Welfare and Research Committee, and reporting complies with ARRIVE guidelines 2.0. Figure 6 shows the holder commissioned for mouse ear irradiations *in vivo*. A soft padded ear clamp containing an 8 mm diameter beam entrance hole was fitted over the left ear of anesthetized mice (injected with ketamine 100 mg/kg xylazine 8 mg/kg IP injection using 24-27g needle) and fixed in a pre-warmed aluminum base. A clear plexiglass top was fitted over the mouse and fixed in place. The proton beam path before and after the ear was clear to allow for *in situ* film dosimetry. The whole assembly was placed on the beamline and radiochromic film was taped behind the ear on the holder to measure the dose delivered to the ear. This design was aimed at minimizing stress to the mice. Mice in the sham cohort were anesthetized and their ears clamped into the holder to assess the effect of clamping on the ear thickness, erythema and desquamation. It was found to be minimal. Animals were excluded from downstream analysis if unforeseen technical issues during irradiation (e.g., alignment deviations) resulted in delivery of a dose that did not match the nominal prescribed dose. In addition, animals that did not survive to the predefined experimental endpoint, for reason unrelated to radiation exposure, were removed from the study.

## Reference mouse irradiations with XRAD

Reference irradiations were carried out with an XRAD320 X-ray source set to 300 kVp. Mice were prepared equally as for LD proton exposures and the same soft padded ear clamp was used to fix the ear in place on a custom-designed holder. A 2 inch by 4 inch by 8 inch lead brick with an 8 mm diameter collimating channel was aligned to the mouse ear and served to protect the remaining mouse body from irradiation. RCF (EBT3) and ion chamber measurements were conducted to establish the dose rate and total dose on the ear, as well as confirming that the rest of the body received no detectable dose.

## Skin reaction post-irradiation observations

Ear thickness was measured longitudinally using a constant-pressure micrometer (Mitutoyo) at baseline (pre-radiation) and at 18 subsequent timepoints spanning up to 85 days post-radiation. Measurements were performed under isoflurane anesthesia. This is a measurement that does not apply pressure to the ear. Left and right ears were measured independently and recorded for each animal. Erythema and desquamation were scored according to four grades, using the same scoring system as in Dombrowsky *et al.*[54]. The final score is the sum of erythema (severe: score 3, definite: score 1.5, mild: score 0.5, no: score 0) and desquamation (moist: score 3, crust: score 2, dry: score 1, no: score 0).

Animals were euthanized under deep isoflurane anesthesia by cervical dislocation and ears from irradiated and sham-exposed mice were harvested using a sterile 8-mm full-thickness biopsy punch. Each punch was bisected with an RNase-free blade; one half was snap-frozen in liquid nitrogen and stored at −80 °C for RNA extraction, and the other half was fixed in 10% neutral-buffered formalin (24-48 h, room temperature) for histological processing.

## RNA sequencing and analysis

RNA was extracted using mechanical disruption followed by purification with the RNAeasy kit (Qiagen). Library preparation and sequencing was performed by the UCLA TCGB Core using a stranded mRNA (poly-A) workflow. Libraries were sequenced on a NovaSeq X Plus (10B, 200-cycle kit). Raw demultiplexed paired-end FASTQ files and sample metadata were provided, verified by checksums (md5/sha256), and stored in a read-only directory to preserve raw data. RNA quality was assessed by BioAnalyzer (RIN from 8.3 to 9.6). RNA-sequencing reads were mapped to the mouse genome (GRCm38/mm10 reference) using align function in Rsubread package (version 2.0.1) with default parameters and per-gene counts of uniquely mapped reads were obtained with featureCounts (v2.0.1). Differential expression analysis was performed using edgeR (version 3.32.1). *p*-values were adjusted for multiple testing using the Benjamini–Hochberg method to control false discovery rate (FDR). Genes with $|\log_2 \text{fold change}| \geq 0.585$ and FDR $\leq 0.05$ were considered significantly differentially expressed (Supplementary Table 2). Quality control included volcano plots (ggplot2). Sample clustering and normalized count heatmaps were generated to confirm reproducibility and identify potential outliers prior to downstream interpretation.

GO Biological Process Enrichment Analysis and Visualization: biological processes of genes with |log2 fold change| ≥ 0.585 and FDR ≤ 0.05 were identified using the clusterProfiler package in R (version 3.18.1) and compared between treatment groups. Venn diagrams were generated with enrichment sets that were filtered to Biological Process ontology and significance by adjusted *p*-value (FDR/padj ≤ 0.05) (falling back to raw $p \leq 0.05$ only if adjusted values were unavailable). Overlaps were computed primarily by GO ID (with per-file deduplication by ID); when IDs were absent, normalized term names were used.

## Statistical Analysis

Ear thickness was measured repeatedly in each animal, so measurements within an animal are not independent, and the animal was treated as the unit of replication. For groups compared against each other quantitatively, group sizes were $n = 8$ (36.5 Gy LD protons, group A), $n = 8$ (35.1 Gy X-rays, group E) and $n = 10$ (40.8 Gy X-rays, group F).

The primary analysis compared groups on one per-animal summary: time-averaged ear thickness, defined as the area under each animal's thickness-versus-time curve (trapezoidal integration of linearly interpolated measurements) divided by the window length. The window was days 18-28 after irradiation, which covers the rise and peak of swelling and is the interval over which all animals were followed, because four of the eight proton-irradiated animals were euthanized on day 28 for planned tissue collection. Animals were required to have measurements spanning the whole window. Per-animal values were compared between groups with a two-sided Welch's t-test and, as a non-parametric check, the Mann-Whitney U test. Group A was compared separately with group E and with group F. Holm adjustment for these two comparisons did not change the conclusions (adjusted Welch $p = 0.048$ for group E and $1.5\times10^{-4}$ for group F); unadjusted p-values are reported in the main text. Sensitivity analyses used the window of days 1-28 and, instead of the time average, each animal's peak thickness (maximum measured value within the window).

The four group A animals euthanized on day 28 did not differ detectably from the four followed to day 81 over days 18-28 (Welch $p = 0.41$, $n = 4$ vs 4), and the two subsets were pooled. The reduction relative to 40.8 Gy X-rays was significant in each subset separately (Welch $p = 5.5\times10^{-5}$ and $8.9\times10^{-4}$). For 35.1 Gy X-rays, $p = 0.026$ for the day-28 subset and $p = 0.13$ for the subset followed to day 81.

Time-resolved comparisons (Fig. 3a) are descriptive. Because the X-ray cohorts were measured on different days than the proton cohort, each X-ray animal's curve was linearly interpolated onto the proton measurement days within the X-ray measurement range, and Welch's t-test was applied at each of these days. *P*-values were adjusted with the Holm procedure for the number of days tested, separately for each comparison. Because the same animals contribute at every time point, these tests are not independent, and they were not used for inference on the overall effect. The number of proton-irradiated animals was 8 until day 28 and 4 thereafter.

As a sensitivity analysis for the comparison with 40.8 Gy X-rays, we fitted a linear mixed-effects model of ear thickness with group, a spline of time and their interaction as fixed effects and a random intercept for each animal. Because a random intercept does not capture between-animal differences in the magnitude and timing of swelling, the model's standard errors are smaller than those of the per-animal analysis, so it was used only as a check on the per-animal result and not for inference.

Additional statistical methods used to evaluate the results of this study are described in the main body of the manuscript where appropriate. Analyses were performed in Python (SciPy 1.11.4, statsmodels 0.15.0). $P < 0.05$ was considered significant.

**Acknowledgments**

The authors gratefully acknowledge the contributions from Samuel Barber, Aodhán McIlvenny, Csaba Tóth, Art Magana, Joe Riley, Zac Eisentraut, Mark Kirkpatrick, Tyler Sipla, Nathan Ybarrolaza, Derrick McGrew, and Teo Maldonado Mancuso.

The work was supported by the U.S. Department of Energy's (DOE) Office of Science (SC) Fusion Energy Sciences (FES) program: the LaserNetUS initiative at the BELLA Center, and the Office of High Energy Physics (HEP), both under Contract No. DE-AC02-05CH11231, as well as the Lawrence Berkeley National Laboratory Laboratory Directed Research and Development (LDRD) funding provided by the Director.

**Author contributions:**

A. M. S., L. O., J. I. and K. N. were the lead scientists of the project.
E. E., B. A. S., C. G. R. G., C. B. S., and J. v. T. supervised the research.
A. J. G. provided scientific support.
J. D. C., S. H., M. C., H. C. and J.-H.M. performed the experiments and data analysis with lead scientists.
L. O. and J. I. wrote the paper, with all authors participating in discussions and data interpretation.

## Figures and Tables

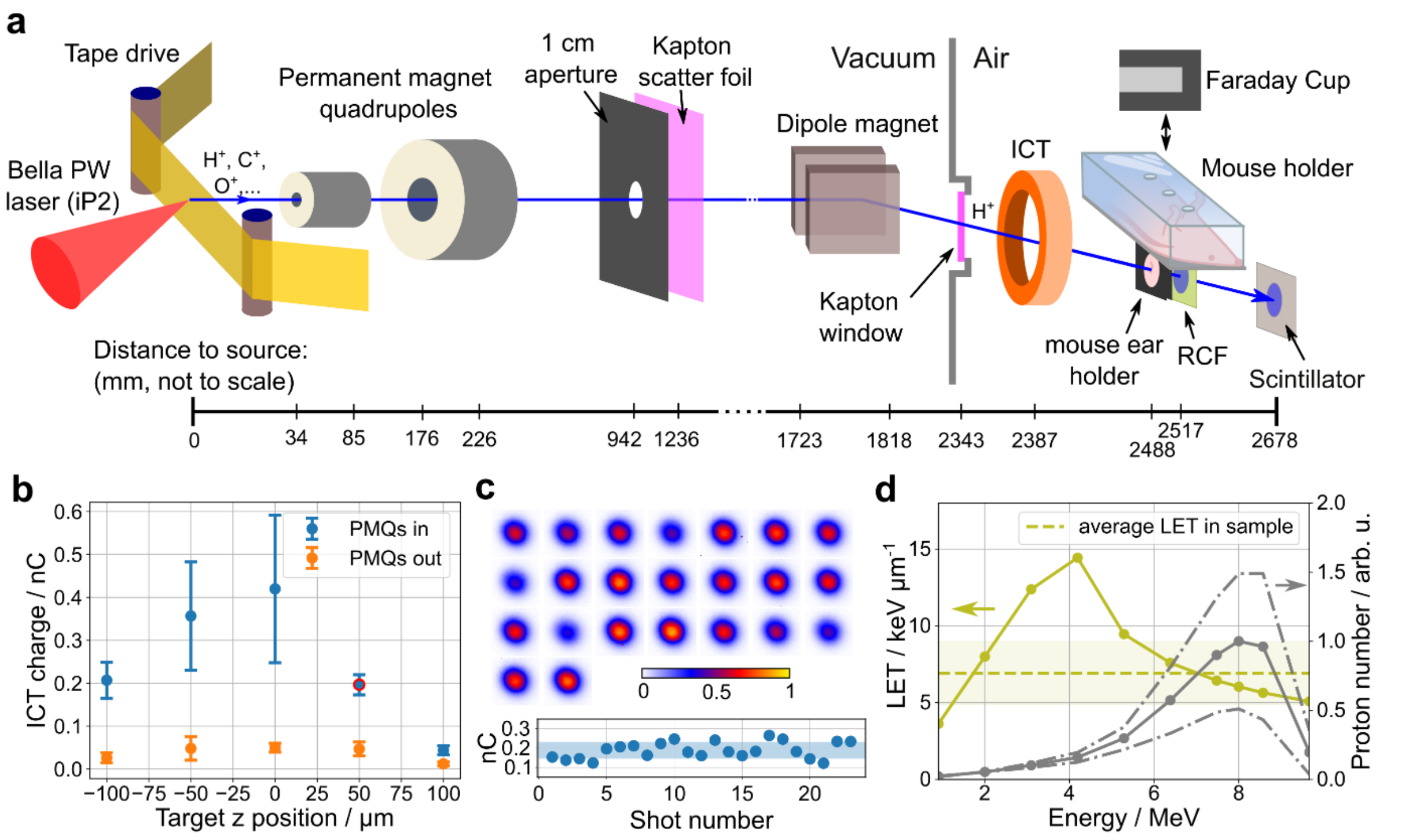

**Figure 1. Laser-driven proton beamline for the generation and delivery of ultra-high instantaneous dose rate proton bunches. a** Schematic depiction of laser-driven proton source, transport beamline, dosimetry, and sample irradiation station at the BELLA PW laser short-focal length setup iP2. Distances and object sizes are not to scale. Details are provided in the main text. **b** Charge measurement at the sample site using an integrating current transformer ("ICT" in a), with (blue) and without (orange) the two permanent magnet quadrupoles inserted into the beam for capture and transport. Markers represent averages over 5-10 laser shots with error bars indicating the standard deviation. Sample exposures were performed with the laser target positioned at a distance z = 50 µm (marked with a red circle) from optimum focus, maximizing the stability of laser-accelerated charge from shot to shot, rather than maximizing the absolute charge delivered to the samples. A negative (positive) target z position corresponds to the target being placed downstream (upstream) of best focus. Position numbers are stage positions, where the stage was oriented along the target normal at a 30° angle with respect to the laser incidence. **c** Example of beam stability over the course of 23 laser shots during sample irradiations as measured with the ICT (bottom, shaded area represents ±1 standard deviation from the average across the data points shown) and a scintillator (top, normalized scintillator light yield) positioned downstream of the samples. **d** Simulated proton spectrum (grey) before entering the sample, with dash-dotted grey lines indicating the range of proton spectra over different exposure days. Linear energy transfer (LET) for protons in a 250 µm thick mouse ear depending on their initial energy (green). The average LET (dashed green line) is weighted with the average proton spectrum. Shaded areas indicate ±1 standard deviation.

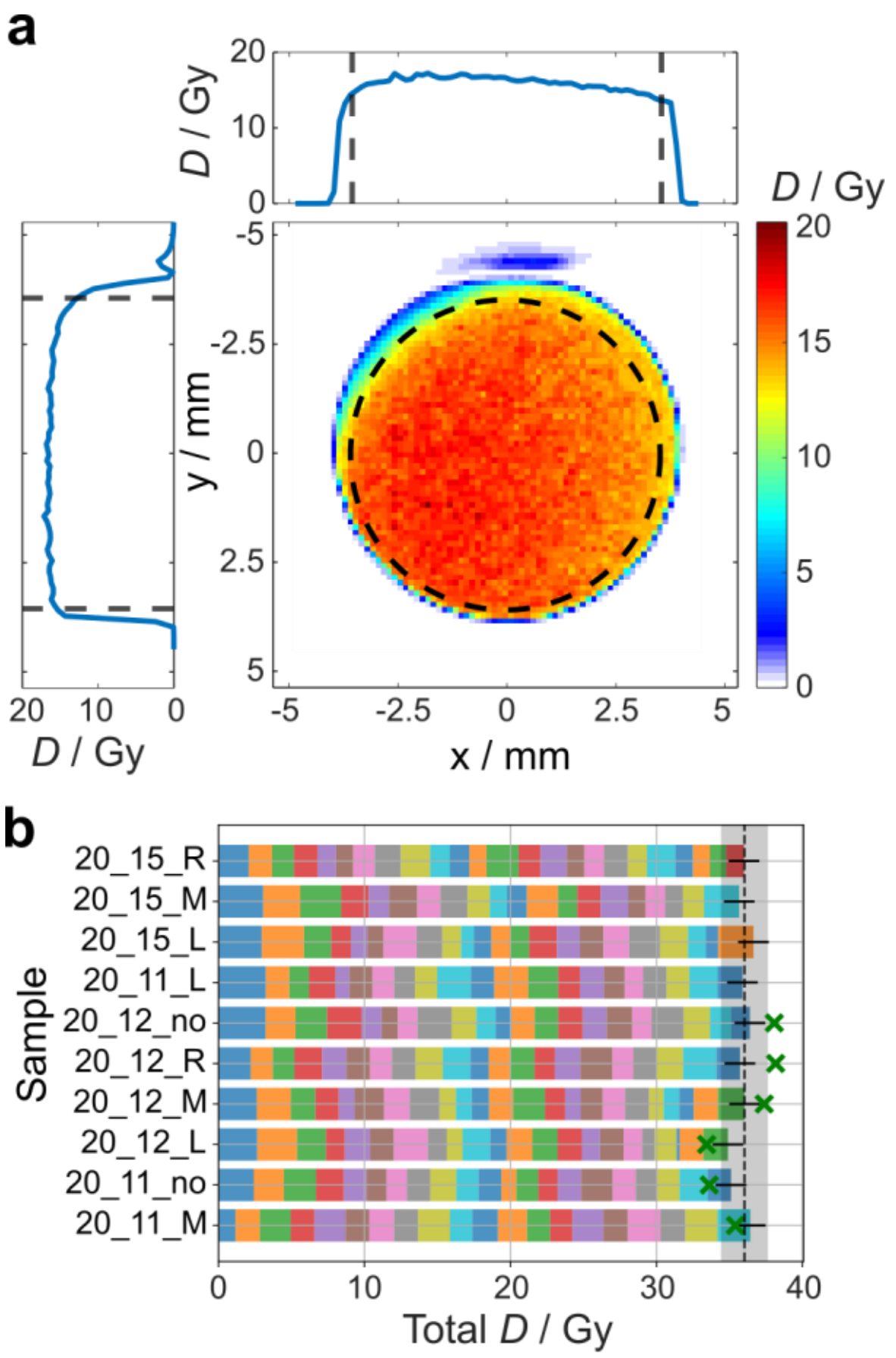

**c**

| Group | *D* / Gy | *SSV* / % | Δ*D* / Gy |
|---|---|---|---|
| A (1f) | 36.5 | 4 | 4.4 |
| B (2f) | 39.6 | 5 | 4.6 |
| C (4f) | 42.1 | 9 | 5.8 |
| D (1f) | 51.0 | 3 | 5.8 |

**Figure 2. Dosimetry results for laser-driven protons**. **a** Proton dose profile at the sample location. The black dashed circle marks the area selected for dose analysis, in this case resulting in an average dose of 15.6 Gy on the film and a lateral dose variation of 6% (standard deviation). Sharp edges result from the aperture of the mouse ear clamp, which serves as a collimator. Horizontal and vertical lineouts are taken at the center of the circle. **b** Accumulated dose applied to samples based on calibrated ICT measurements for group A (36.5 Gy applied in one fraction). Each colored block corresponds to the dose applied in one laser shot. Green crosses represent the dose delivered to the sample, as determined from dosimetry measurements using the RCF placed behind the sample. **c** Table of dosimetry results for each mouse group. Samples in group A (B, C, D) received an average total dose $D$ = 36.5 (39.6, 42.1, 51.0) Gy in 1 (2, 4, 1) fraction. The sample to sample variation of $D$ ranged from 2 % (group D) to 9 % (group C). The total dose error $\Delta D$ was between 4.4 Gy (group A) and 5.8 Gy (group C, D).

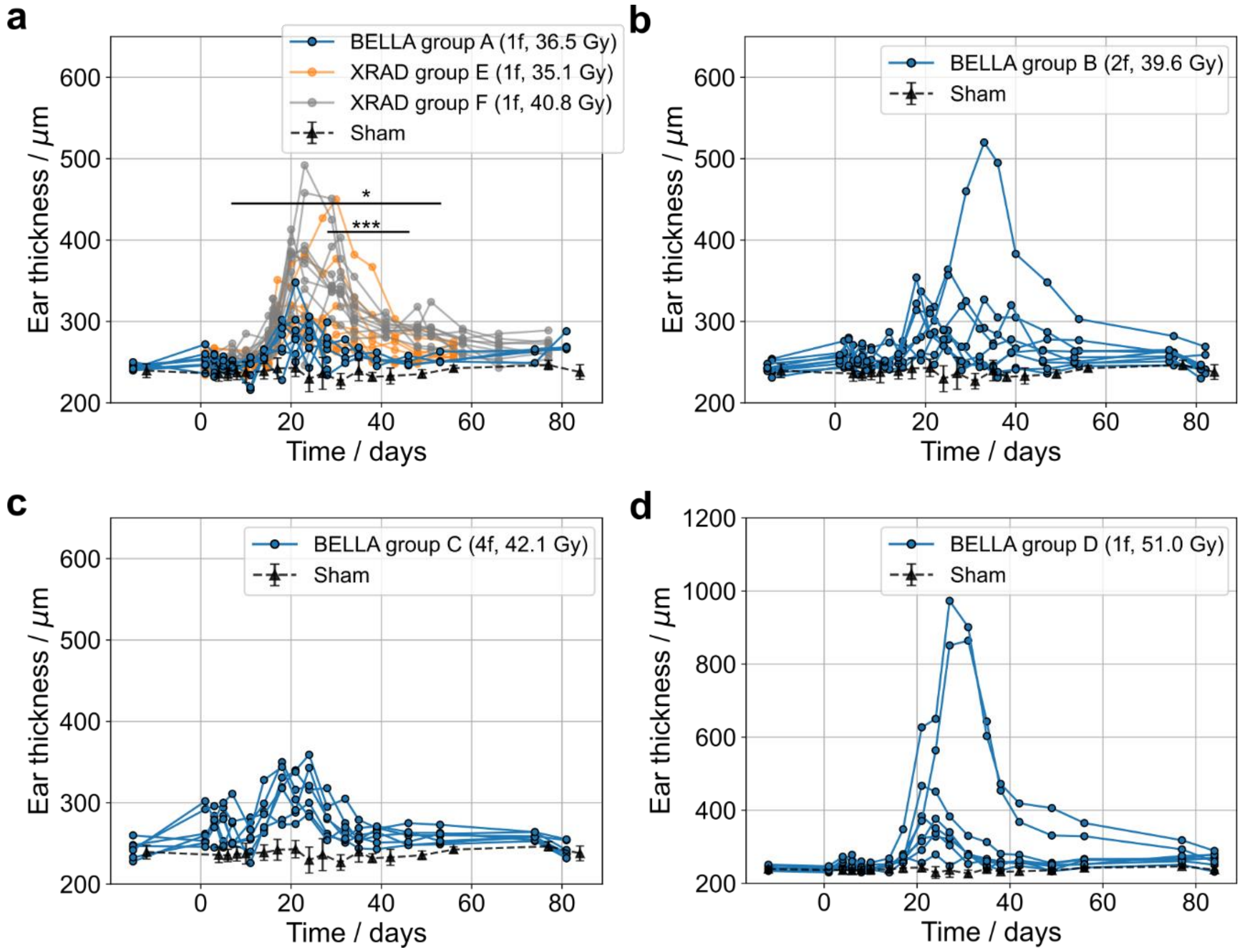

**Figure 3**. **Development of ear thickness after the (last) day of irradiation, taking place on day 0**. **a** Mouse ears irradiated with 36.5 Gy of BELLA protons within the same day (group A, blue), 35.1 Gy X-rays (group E, orange), and 40.8 Gy X-rays (group F, grey). Horizontal lines indicate at what times the Holm-adjusted Welch *p*-value comparing 36.5 Gy BELLA protons to 40.8 Gy X-rays is $p \leq 0.05$ (*) and $p \leq 0.001$ (***). X-ray values were linearly interpolated onto the proton measurement days to calculate *p*-values, which are descriptive only due to non-independence of subsequent measurements. The proton group contained $n = 8$ animals until day 28 and $n = 4$ animals thereafter. **b** Mouse ears irradiated with a total dose of 39.6 Gy of BELLA protons over two subsequent days (group B), **c** 42.1 Gy over four subsequent days (group C), and **d** 51.0 Gy within the same day (group D). Black triangles indicate the ear thickness of the BELLA proton sham cohort over the same time period. The XRAD sham cohorts did not differ significantly from the proton sham cohort (Supplementary Figure 3).

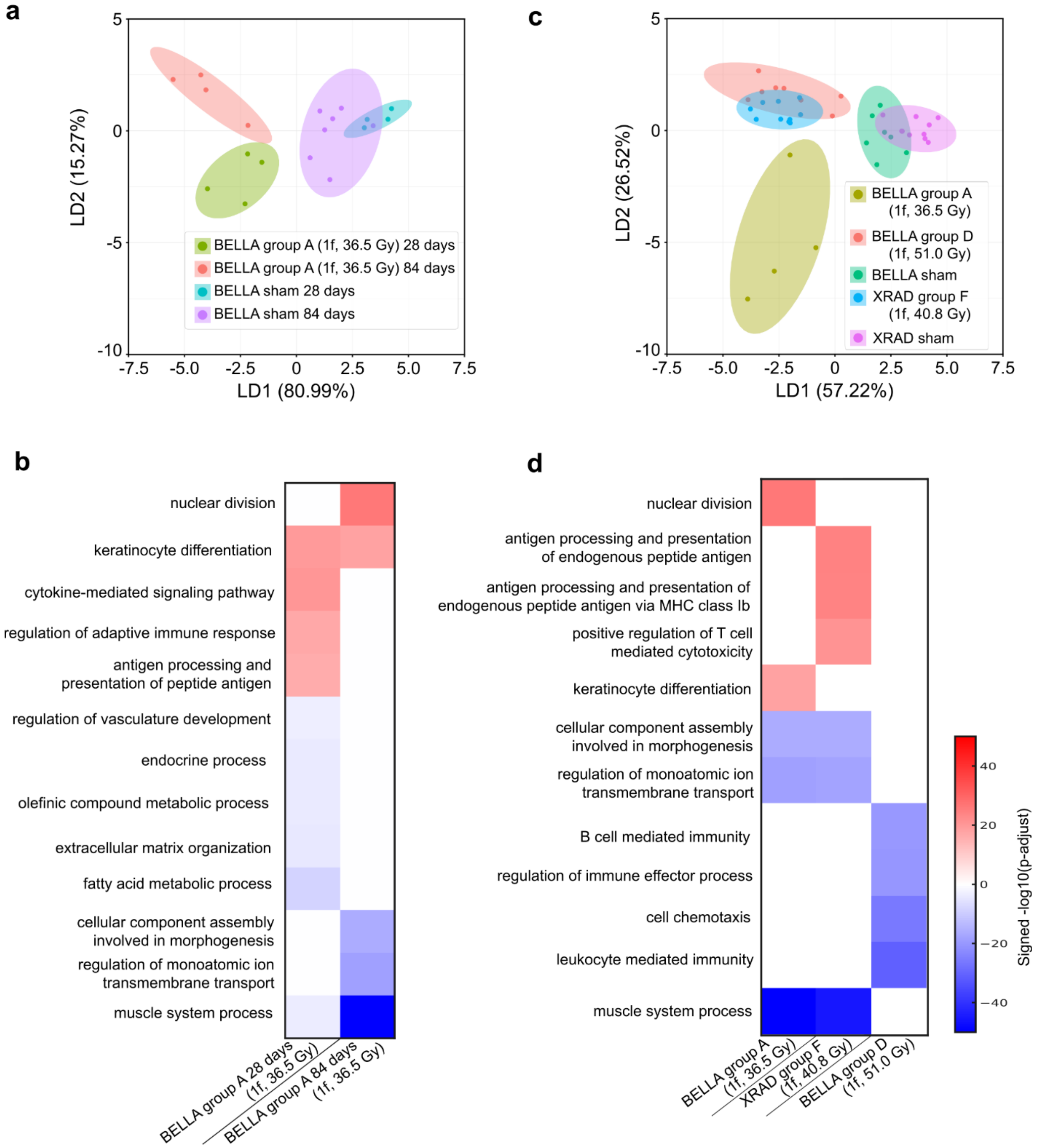

**Figure 4**. **Transcriptional responses to LD proton irradiation are dose- and time-dependent. a** LDA of RNA-seq data from BELLA Group A (36.5 Gy) shows clear separation between 28-day and 84-day irradiated transcriptomes, with both distinct from their respective shams; 28-day and 84-day sham controls overlap, indicating stable baselines. **b** GO Biological Process enrichment heatmap comparing 28-day and 84-day responses to 36.5 Gy BELLA. Thirteen representative processes spanning immune, keratinocyte, metabolic, ECM/structural, and muscle-system processes are shown as signed $-\log10(p.\text{adjust})$ values (red = enriched among upregulated genes; blue = enriched among downregulated genes; white = no significant enrichment). Early cytokine, antigen-presentation, and keratinocyte-associated enrichments present at 28 days are absent by 84 days, while enrichments among downregulated genes shift toward muscle-system, ECM-organization, and morphogenesis-related processes. Keratinocyte-differentiation and proliferative gene enrichments (e.g., nuclear division) remain detectable at 84 days. Significance was determined by Benjamini–Hochberg–adjusted *p*-values; only GO terms with FDR < 0.05 were considered significant. **c** LDA of late (84-day) transcriptomes across modalities shows dose-dependent divergence between BELLA Group A (36.5 Gy) and BELLA

Group D (51.0 Gy), whereas BELLA Group D partially overlaps with XRAD Group F (40.8 Gy), suggesting shared late transcriptional features at higher doses. Shams cluster tightly and remain separate from all irradiated groups. **d** Signed GO enrichment heatmap of the top late-time (84-day) Biological Processes across BELLA Group A (36.5 Gy), XRAD group F (40.8 Gy) and BELLA Group D (51.0 Gy). All displayed non-zero GO terms in **b** and **d** have FDR<0.001.

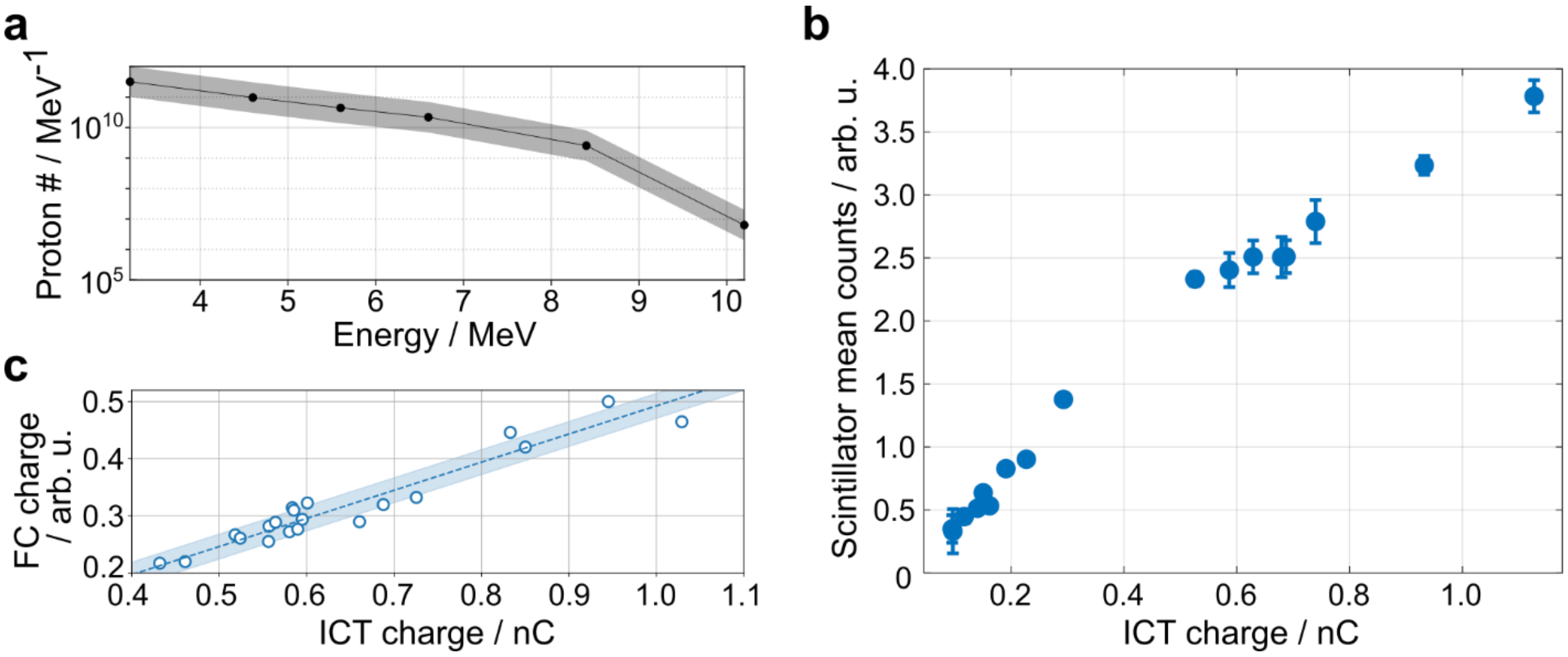

**Figure 5**: **Proton diagnostics results. a** TNSA source spectrum measured with a radiochromic film stack at 52 mm behind the TNSA target. The shaded area shows an uncertainty estimate of ~0.5 orders of magnitude which is typical for this type of measurement[79]. **b** Scintillator signal (averaged within an area matched to the sample size and centered at the peak location of the beam, error bars indicate the standard deviation) in comparison with ICT measured charge over the course of 17 laser shots during beam tuning while varying the laser intensity on target. The Pearson correlation coefficient of $p$ = 0.99 between these two measurements indicates a strong correlation. **c** Charge measured with a Faraday Cup (FC) and simultaneously measured with the ICT. The dashed line shows a linear regression and the shaded area represents the standard deviation of residuals from the linear regression.

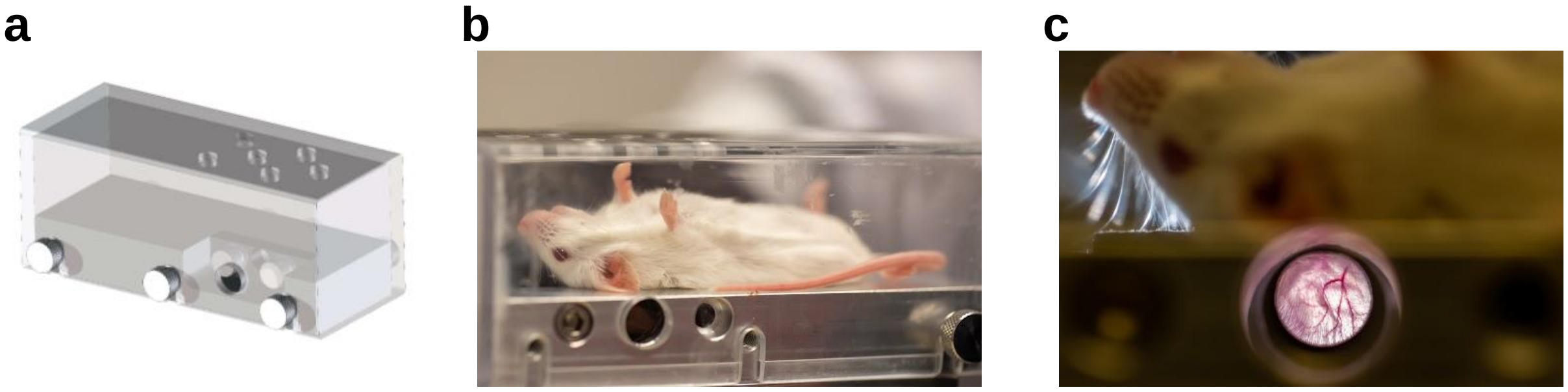

**Figure 6. Mouse ear irradiation assembly. a** Design of the assembly with ear clamp, aluminum base and plexiglass top with breathing holes. **b** Photographs of the mouse holder with anesthetized mouse and clearing for the proton beam path during irradiation campaign at the BELLA PW proton beamline. **c** Close-up of the proton beam path and mouse ear, showing vascular structure.

| Proton energy | LET | Dose per bunch | Bunch length | Bunch separation | IDR | MDR |
|---|---|---|---|---|---|---|
| 8 MeV | 6.9 keV/µm | 2 Gy | 11 ns | ~20 s | $1.3 \times 10^8$ Gy/s | ~0.1 Gy/s |

**Table 1. Summary of proton beam parameters for sample exposures at the LD proton beamline of the BELLA PW. Here, proton energy denotes the peak of the on-sample energy distribution (energy distribution range: 5–9 MeV).**